**Effect of damage by 2-MeV He ions and annealing on $H_{c2}$ in MgB$_2$ thin films**

R. Gandikota,[a)] R.K. Singh,[a)] J. Kim,[a)] B. Wilkens,[b)] N. Newman,[a),b),g)] J.M. Rowell,[a)] A.V. Pogrebnyakov,[c),d),e)] X.X. Xi,[c),d),e)] J.M. Redwing,[d),e)] S.Y. Xu,[c),e)] Qi Li,[c),e)] and B.H. Moeckly[f)]

[a)] Department of Chemical and Materials Engineering, Arizona State University, Tempe, Arizona 85287-6006.

[b)] Center for Solid State Science, Arizona State University, Tempe, Arizona 85287-1704.

[c)] Department of Physics, The Pennsylvania State University, University Park, Pennsylvania 16802.

[d)] Department of Materials Science and Engineering, The Pennsylvania State University, University Park, Pennsylvania 16802.

[e)] Materials Research Institute, Pennsylvania State University, University Park, Pennsylvania 16802.

[f)] Superconductor Technologies, Inc., 460 Ward Drive, Santa Barbara, CA 93111

[g)] Author to whom correspondence should be addressed; Electronic mail: Nathan.Newman@asu.edu

**Abstract**

The effect of damage induced by 2-MeV alpha particles, followed by annealing, on the critical temperature ($T_c$), resistivity ($\rho$), and upper critical field ($H_{c2}$), of three $MgB_2$ films made by different deposition processes has been studied. Damage creates a linear decrease in $T_c$ with residual resistivity ($\rho_0$), and produces maxima in both $H_{c2}^{\perp}(0)$ and $H_{c2}^{\parallel}(0)$. Below $T_c$s of about 25 K, $H_{c2}(0)$ depends roughly linearly on $T_c$, while the anisotropy of $H_{c2}(0)$ decreases as $T_c$ decreases. Annealing the films reproduces the $T_c$ vs. $\rho_0$ dependence but not the $H_{c2}(0)$ values induced by damage.

Carrier scattering within and between the three dimensional π-band and two-dimensional σ-band of carriers in superconducting $MgB_2$ leads to large variations in the transition temperature ($T_c$), resistivity (ρ) and upper critical field ($H_{c2}$). Of particular interest is the large increase in $H_{c2}$ that results from alloying with carbon,[1-4] and also, in the case of thin films, from disorder and possibly impurities that are introduced during deposition.[5] It has been shown that in $MgB_2$ there is no correlation of $H_{c2}(0)$ values with resistivity,[6] which can be explained most simply if $H_{c2}(0)$ is determined by scattering in the low conductivity σ-band, while the resistivity is determined by scattering in the high conductivity π-band. Films alloyed with carbon exhibit $H_{c2}$ values exceeding those of NbTi and $Nb_3Sn$,[5] making $MgB_2$ a candidate material for many high field applications, while allowing operation at 20 to 25 K. We find that ion damage and annealing of thin films which were deposited in three different ways results in changes in $H_{c2}(0)$, ρ and $T_c$ that can only be explained if the intraband scattering processes are being modified independently.

Three $MgB_2$ thin films (A, S, and PB), all grown on sapphire substrates, were used for this study, while PA was used earlier.[7] Films PA and PB were grown at Pennsylvania State University (PSU), film S at Superconductor Technologies Inc. (STI), and film A at Arizona State University (ASU) using methods described previously.[8-10] The deposition temperatures and physical properties of the films are given in Table 1. A high resistivity $Ta_xN$ film was deposited as a passivation layer and the films were patterned and ion irradiated as before.[7] Measurements of ρ(T) and ρ(H) were made before and after each irradiation step using a 9 T Quantum Design Physical Property

Measurement System (PPMS). $H_{c2}$ was defined as $\rho(H_{c2})=0.9\rho_0$. Annealing was carried out in $N_2$ atmosphere.

After the $T_c$ of film PB was reduced to 21.3 K, cracking of the film prevented any further measurements. The $T_c$s of films S and A were suppressed to about 6 K after irradiation to ion fluxes of 1.8 x$10^{17}$ /cm$^2$ and 3.4x$10^{17}$ /cm$^2$ respectively. We reported earlier that for film PA, as the irradiation dose increased, $T_c$ decreased linearly with the intragrain resistivity measured at 40 K, while the intergrain connectivity was little changed.[7] The behaviors of the three films reported here are very similar. After correcting the measured resistivity of the films for limited connectivity[11,12] (which is significant only in film A), the obtained corrected residual resistivity ($\rho_{0,corr}$) when extrapolated to $T_c$=0 K falls between 70 μΩ cm and 93 μΩ cm (Table 1) for the four films. Similar plots of $T_c$ vs. $\rho_{0,corr}$ using the data available for carbon-alloyed[13] and neutron-damaged bulk samples,[14] show that the extrapolation to $T_c$=0 K occurs at 51.1 μΩ cm and 106.5 μΩ cm respectively.

Fig.1 shows $H_{c2}^{\parallel}(0)$ and $H_{c2}^{\perp}(0)$ for film S before irradiation and after irradiation steps have reduced $T_c$. For clarity, only 8 of the measurements made at 14 $T_c$ values have been shown. As $T_c$ is reduced, it can be seen that $H_{c2}(0)$ for both field orientations initially increases, but as damage continues, the further decrease in $T_c$ is accompanied by a decrease in $H_{c2}(0)$, with $dH_{c2}^{\parallel}/dT(T_c)$ remaining roughly constant from 31 K to 24 K at a value of 0.57 T/K. Similar behavior has been observed for neutron damage in bulk $MgB_2$.[14]

As the magnetic field is limited to 9 T, determination of $H_{c2}(0)$ requires extrapolation to higher fields at $T$=0 K. For films with $H_c$ linear with $T$ from 4 T to 9 T, $H_{c2}(0)$ was determined by a linear extrapolation of this dependence to $T=0$ K. A parabolic extrapolation of the downward curvature of $H_c(T)$ was used in films with lower $T_c$s. Such extrapolations might be in error by 20 to 30%, underestimating the true value[5] in cases where $H_{c2}(T)$ has an upturn at higher magnetic fields than 9 T.

The $H_{c2}(0)$ values of the films before damage are given in Table 1. The spread in values arises likely from the differences in disorder and scattering resulting from the deposition processes, with the highest $H_{c2}$s being seen in film A, which has the highest resistivity. The anisotropy of $H_{c2}(0)$ ($\gamma$) across the three samples (Table 1) possibly reflects the degree of orientation of the films. TEM studies of other ASU films indicate a very small grain size, with limited crystallinity and limited c-axis orientation. Other STI films exhibit both energy gaps in point contact tunneling studies, consistent with some mixed orientation. PSU films are reported to be well oriented.[8]

With increasing ion damage, the variations of $H_{c2}(0)$ vs. $T_c$ are also quite different in film A (Fig. 2). The two low resistivity films (S and PB) exhibit maxima in $H_{c2}^{\parallel}(0K)$ of 29 and 34 T respectively at $T_c$s between 30 K and 35 K, and in $H_{c2}^{\perp}(0)$ at 17 and 15 T respectively at $T_c$s just below 30 K. In film A, the initially high $H_{c2}(0)$ values decrease monotonically.

After the $T_c$s of films A and S had been reduced to below 10 K, they were annealed at temperatures from 100 to 350 °C, which increased $T_c$ and decreased the resistivity. As seen in the inset of Fig. 3, the values of $T_c$ and $\rho_{0,corr}$ for film A after annealing fit on the $T_c$ vs. $\rho_{0,corr}$ dependence induced by damage. For film S, the $T_c$s after longer anneals fall a little below the initial dependence. Interestingly, after films A and S had been annealed to 300 °C, the $T_c$s of both were about 24 K.

Measurement of $H_{c2}$ after annealing produced the surprising results shown in Figs. 2 and 3. In film A, annealing results in $H_{c2}(0)$ values that are very close to those that resulted from damage (Fig. 2). However, the $H_{c2}(0)$ values of film S after annealing are lower by almost 40% than those produced by damage alone, even though the $T_c$ and $\rho_0$ values are remarkably similar. Damage increases, and annealing reduces, the width of the transitions in field, although the widths remain larger than those of the as-made films. The widths are particularly large in the high resistivity ASU film (A).

In MgB$_2$ films, scattering can be readily introduced during growth, or it can be induced by damage after deposition, and Fig. 2 illustrates the importance of both types of disorder. Films S and PB initially have high $T_c$s and low resistivities. They also have relatively low values of $H_{c2}$, particularly for $H_{c2}^\perp$, 12 and 6 T respectively. However film A, with $T_c$ of 34.2 K and $\rho_0$ of 89.2 μΩ cm, has $H_{c2}^\perp$ of 26 T, over a factor of two higher.

The effect of scattering induced by ion damage is also film dependent. Films S and PB behave very similarly, with $H_{c2}^{||}$ values reaching broad maxima when $T_c$ is 30 to 35 K, while $H_{c2}^\perp$ is a maximum just below 30 K. In film A, in-situ disorder alone results in a value of $H_{c2}^{||}$ which is a little higher than the maximum induced by damage in films S and PB, while the in-situ $H_{c2}^\perp$ far exceeds the maxima induced by damage in films S and PB. As noted earlier, this high value of $H_{c2}^\perp$ might be a result of limited orientation. In film A, damage results only in a decrease in $H_{c2}$ values. As damage proceeds further, $H_{c2}(0)$ for both field orientations in films A and S decreases approximately linearly with $T_c$ for $T_c$s below about 25 K. The maximum values of $H_{c2}^{||}$ near 35 T, resulting from either in-situ or ion damage induced scattering, are very comparable to the maximum in neutron damaged bulk samples[14] (about 30 to 35 T at a $T_c$ of 36 K). Also, the maximum $H_{c2}$ in bulk carbon alloy samples, including single crystals, is again 35 T, when $T_c$ is 34 to 36 K.[2] It has been suggested that the increase in $H_{c2}$ on alloying with carbon, compared to the decrease with Al, is an effect of scattering.[2] Thus, scattering induced by growth (film A), by ion damage (films S and PB), by neutron damage,[14] and by alloying with carbon results in the same $H_{c2}^{\parallel}$ values, within the errors of extrapolations. It is intriguing that these four means of introducing scattering all result in the same $H_{c2}^{\parallel}(0)$ value as if there is a common upper limit to scattering that can be achieved, before $T_c$ is reduced too much. However, the maxima in $H_{c2}^{\perp}$ in Fig. 2 (26, 17 and 15 T) are much larger than the 7 T value seen in C-alloyed single crystals,[3] possibly due to limited orientation in film A, but this is an unlikely explanation of the film PB value. In carbon-alloyed films,[5] $H_{c2}$ in both orientations is considerably higher (51 T and 35 T measured at 4.2 K) than in any other film or bulk $MgB_2$ samples.

Somewhat divergent views of the relative importance of π and σ intraband scattering can be found in the literature. The fact that there is no correlation of $H_{c2}(0)$ with ρ across numerous samples leads to the suggestion that the measured resistivity is $\rho_\pi$, while $H_{c2}(0)$ is determined by σ-band scattering[15] ($\rho_\sigma$), with the magnitude of $\rho_\sigma$ being 125 to 165 μΩ cm.[6] However, to explain the observed form of $H_{c2}(T)$,[16] and of $\gamma(T)$,[2] π-scattering was considered important.

Reductions in $T_c$ in $MgB_2$ have been widely ascribed to interband scattering, but as the linear $T_c$ vs $\rho_0$ dependence of $MgB_2$, which results from ion irradiation,[7] neutron irradiation,[14] and alloying with carbon,[4] is similar to that seen in many A15 and ternary boride superconductors,[17] it is not necessary to invoke interband scattering to explain such a reduction in $T_c$. Smearing of the peak, and reduction of the density of states at the Fermi level,[18] is an alternative explanation. While $\rho_0$ increases markedly with damage, the form of $\rho(T)$ and value of $d\rho/dT$ at 300 K (about 0.065 $\mu\Omega$ cm/K) are consistent with π-band scattering and change very little, even down to $T_c$s of about 10 K. It is tempting to claim that the measured resistivity in the four films reported here, is $\rho_\pi$, over the whole range of $T_c$s. If the magnitude of $H_{c2}$ depends only on σ band scattering, or $\rho_\sigma$, then the complex behavior of the three films in Fig. 2 is ascribed to changes in $\rho_\sigma$ and $T_c$ induced by ion damage and by annealing. The increase of $H_{c2}(0)$ to the common maximum value near 35 T is then due to an initial increase in $\rho_\sigma$. For lower $T_c$s, the roughly linear dependence of $H_{c2}(0)$ on $T_c$ seen in Fig. 2 suggests that $\rho_\sigma$ reaches a maximum value, which is not changed by further damage, and $H_{c2}$ is then proportional to $T_c$.

The work at ASU was supported by ONR under grant N00014-02-1-0002. We acknowledge use of facilities in the Center for Solid State Science at ASU. The work at Penn State was supported by ONR under grant Nos. N00014-00-1-0294 (Xi) and N0014-01-1-0006 (Redwing), and by NSF under grant Nos. DMR-0405502 (Li), and DMR-0306746 (Xi and Redwing).

[11]It has been pointed out[12] that the limited connectivity of many MgB$_2$ samples results in an increase in $\Delta\rho_{300\text{-}41K}$ [$=\rho(300K)- \rho(41K)$] over the value seen in fully connected samples. The typical value of $\Delta\rho_{300\text{-}41K}$ in a number of PSU and STI films is about 8 $\mu\Omega$ cm (7.4 $\mu\Omega$ cm for the film PB), hence the connectivity factor used in this work is 7.4/$\Delta\rho_{300\text{-}41K}$ (measured), i.e., the intragrain resistivity or $\rho_{corr}(T)$ is $\rho(T)$ x 7.4/$\Delta\rho_{300\text{-}41K}$.

[12]J.M. Rowell, Superconductor science and Technology **16**, 17 (2003).

[13]S. Lee, T. Masui, A. Yamamoto, H. Uchiyama, S. Tajima, Physica C **397,** 7 (2003).

[14]M. Putti, V. Braccini, C. Ferdeghini, F. Gatti, P. Manfrinetti, D. Marre, A. Palenzona, I. Pallecchi, C. Tarantini, I. Sheikin, H.U. Aebersold, E. Lehmann, to be published (APL).

[15]S.L. Bud'ko, V.G. Kogan, and P.C. Canfield, Phys. Rev. B **64**, 180506 (2001).

[16]A. Gurevich, Phys. Rev. B **67**, 184515 (2003).

[17]J.M. Rowell and R.C. Dynes, unpublished.

[18]A.P. Gerashenko, K.N. Mikhalev, S.V. Verkhovskii, A.E. Karkin, and B.N. Goshchitskii, Phys. Rev. B **65**, 132506 (2002).

**List of tables and figures:**

Table 1. Summary of electrical properties exhibited by the films.

Fig. 1 Variation of (a) $H_{c2}^{\parallel}$ and (b) $H_{c2}^{\perp}$ of film S with temperature after selected irradiation steps.

Fig. 2 Variation of $H_{c2}(0)$ with $T_c$ for films A, S, and PB. $H_{c2}$ vs. $T$ curves were extrapolated to 0 K (procedure explained in text) to obtain $H_{c2}(0)$.

Fig. 3 $H_{c2}$ vs. $T$ for similar $T_c$s obtained during irradiation and then later by annealing for (a) film A and (b) film S. The inset in (a) shows the $T_c$ vs. $\rho_{0,corr}$ plot obtained for the two above mentioned films.

| Sample | Deposition Temperature (°C) | $T_c$(K) | $\rho_0$ (as-made) ($\mu\Omega$ cm) | $\Delta\rho_{300\text{-}41K}$= [$\rho$(300K)-$\rho$(41K)] ($\mu\Omega$ cm) | $\rho_{0,corr}$ at $T_c$=0 K ($\mu\Omega$ cm) | Critical fields $H_{c2}^{\parallel}(0)$ (T) | $H_{c2}^{\perp}(0)$ (T) | $\gamma(0)$= $H_{c2}^{\parallel}(0)/H_{c2}^{\perp}(0)$ |
|---|---|---|---|---|---|---|---|---|
| PA | 750 | 40.5 | 1.85 | 11.1 | 78.0 | - | - | - |
| PB | 750 | 41 | 0.51 | 7.4 | 72.3 | 27.5 | 5.9 | 4.66 |
| S | 550 | 38 | 4.3 | 6.5 | 70.7 | 23.1 | 11.5 | 2.01 |
| A | 300 | 34.2 | 89.2 | 36.4 | 92.8 | 34.7 | 25.9 | 1.34 |

**Table 1-Gandikota *et al*.**

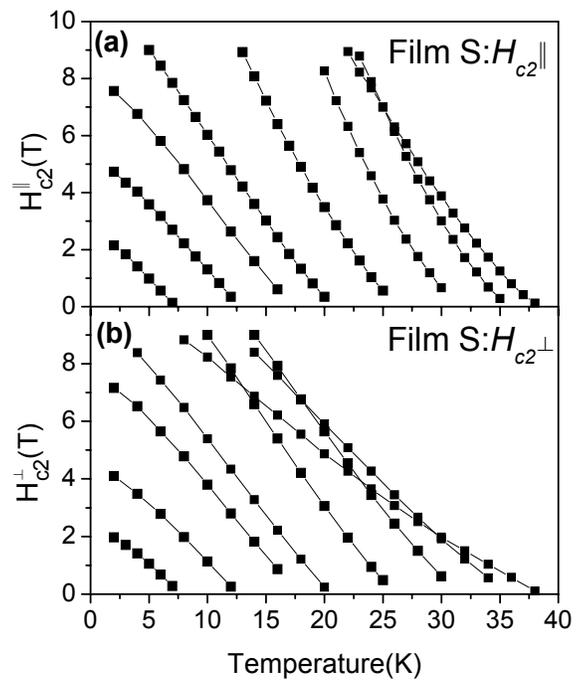

**Fig. 1-Gandikota *et al*.**

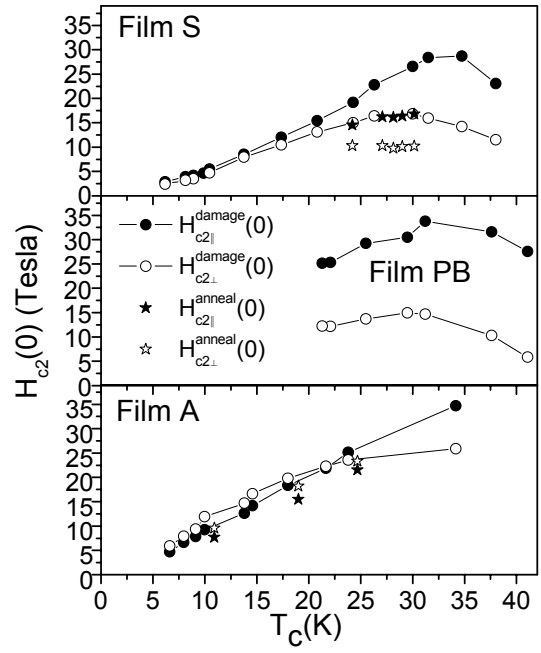

**Fig. 2-Gandikota *et al*.**

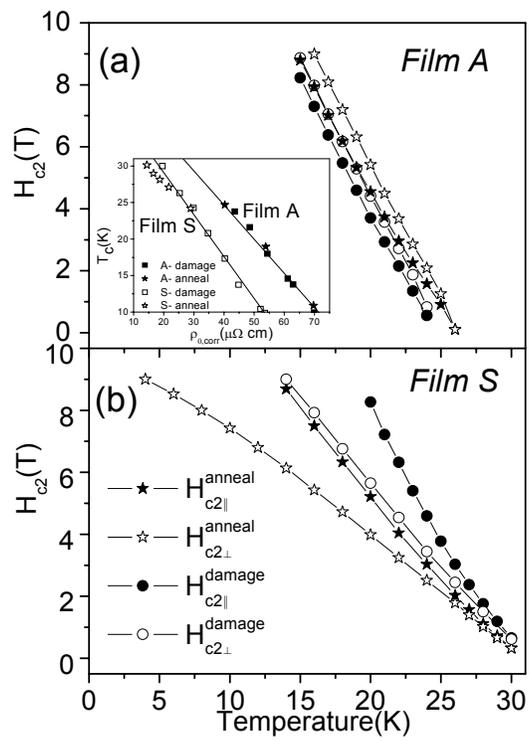

**Fig. 3-Gandikota *et al*.**

15